\documentclass[runningheads,a4paper]{llncs}

\usepackage{amssymb}
\usepackage{amsmath}
\usepackage{graphicx}
\usepackage{color}
\usepackage[table,xcdraw]{xcolor}
\usepackage{url}
\usepackage{multirow}
\usepackage{hyperref}
\usepackage{subfig}
\usepackage{todonotes}
\usepackage[misc]{ifsym}

\urldef{\mail}\path|j.m.wolterink@umcutrecht.nl|

\newcommand{\keywords}[1]{\par\addvspace\baselineskip
\noindent\keywordname\enspace\ignorespaces#1}

\usepackage{array}
\newcolumntype{L}[1]{>{\raggedright\let\newline\\\arraybackslash\hspace{0pt}}m{#1}}
\newcolumntype{C}[1]{>{\centering\let\newline\\\arraybackslash\hspace{0pt}}m{#1}}
\newcolumntype{R}[1]{>{\raggedleft\let\newline\\\arraybackslash\hspace{0pt}}m{#1}}

\begin{document}

\mainmatter  

\title{Graph Convolutional Networks \\ for  Coronary Artery Segmentation \\ in Cardiac CT Angiography}
\titlerunning{Coronary Artery Segmentation using Graph Convolutional Networks}

\author{Jelmer M. Wolterink\inst{1} \and Tim Leiner \inst{2} \and Ivana I\v{s}gum\inst{1}}


\authorrunning{J.M. Wolterink et al.}

\institute{Image Sciences Institute, University Medical Center Utrecht, The Netherlands \and Department of Radiology, University Medical Center Utrecht, The Netherlands}
\maketitle

\begin{abstract}
Detection of coronary artery stenosis in coronary CT angiography (CCTA) requires highly personalized surface meshes enclosing the coronary lumen. In this work, we propose to use graph convolutional networks (GCNs) to predict the spatial location of vertices in a tubular surface mesh that segments the coronary artery lumen. Predictions for individual vertex locations are based on local image features as well as on features of neighboring vertices in the mesh graph. The method was trained and evaluated using the publicly available Coronary Artery Stenoses Detection and Quantification Evaluation Framework. Surface meshes enclosing the full coronary artery tree were automatically extracted. A quantitative evaluation on 78 coronary artery segments showed that these meshes corresponded closely to reference annotations, with a Dice similarity coefficient of 0.75/0.73, a mean surface distance of 0.25/0.28 mm, and a Hausdorff distance of 1.53/1.86 mm in healthy/diseased vessel segments. The results showed that inclusion of mesh information in a GCN improves segmentation overlap and accuracy over a baseline model without interaction on the mesh. The results indicate that GCNs allow efficient extraction of coronary artery surface meshes and that the use of GCNs leads to regular and more accurate meshes.
\end{abstract}

\keywords{Graph convolutional networks, coronary CT angiography, coronary arteries, lumen segmentation}

\section{Introduction}
Coronary CT angiography (CCTA) images provide valuable information to determine the anatomical or functional severity of coronary artery stenosis \cite{Leip14}. Methods for stenosis detection \cite{Kiri13} and blood flow simulation \cite{Tayl13} typically require highly personalized coronary lumen surface meshes with sub-voxel accuracy. Because manual segmentation of the full coronary artery tree in a CCTA image would hardly be feasible, such meshes are typically extracted using automatic or semi-automatic methods \cite{Lesa09,Luga14,Frei17}. Deep learning-based segmentation could further improve such methods \cite{Litj17}, but widely adopted voxel-based segmentation methods do not meet the requirements of down-stream applications, i.e. sub-voxel accuracy and segmentation of the lumen as a contiguous structure. 

An alternative to voxel segmentation is to incorporate a shape prior, i.e. to exploit the fact that an individual vessel or vessel segment has a roughly tubular shape. The wall of this tube can be deformed to match the visible lumen in the CCTA image and obtain an accurate segmentation \cite{Lee19}. In one variant of this approach, a tubular surface mesh along the coronary artery centerline is considered and the spatial location of each mesh vertex is predicted so that the surface closely follows the artery wall.  
Lugauer et al. \cite{Luga14} used probabilistic boosting trees to predict these locations based on steerable image texture features extracted from 2D cross-sectional images. As predictions were made based on 2D cross-sectional information, additional smoothing of the surface mesh was required through graph-cut optimization. Freiman et al. used a similar approach, but enforced smoothness of the obtained surface by penalizing jumps between neighboring vertices \cite{Frei17}. 

In this work, we propose to directly optimize the location of the tubular surface mesh vertices using graph convolutional networks (GCNs) \cite{Kipf16,Hami17}. GCNs are a recent development in deep learning-based medical image analysis, with high potential for graph-based applications in e.g. airway extraction in chest CT \cite{Selv18} and cortical segmentation in brain MR \cite{Cucu18}. We consider the vertices on the coronary lumen surface mesh as graph nodes and solve a regression problem for each of these graph nodes. Predictions for vertices depend on local features as well as on internal representations of adjacent vertices on the mesh. Our experiments show that GCNs are well-suited to accurately predict vertex locations for this natural graph representations, and that direct operation on the tubular mesh reduces the need for additional post-processing.

\section{Data}
We included coronary CT angiography (CCTA) images from the training set of the Coronary Artery Stenoses Detection and Quantification Evaluation Framework \cite{Kiri13}. This set contains 18 CCTA images acquired on Philips, Siemens, and Toshiba CT scanners. Images had an in-plane resolution of 0.29--0.43 mm$^2$ and a slice spacing of 0.25--0.45 mm. Within the challenge, reference surface meshes were annotated by three observers in a selection of segments in each data set, for a total of 78 coronary artery segments. For each CCTA volume, we automatically extracted centerlines in all volumes using our previously proposed deep learning-based method \cite{Wolt19}.

\begin{figure}[t!]
    \centering
    \includegraphics[width=0.8\linewidth]{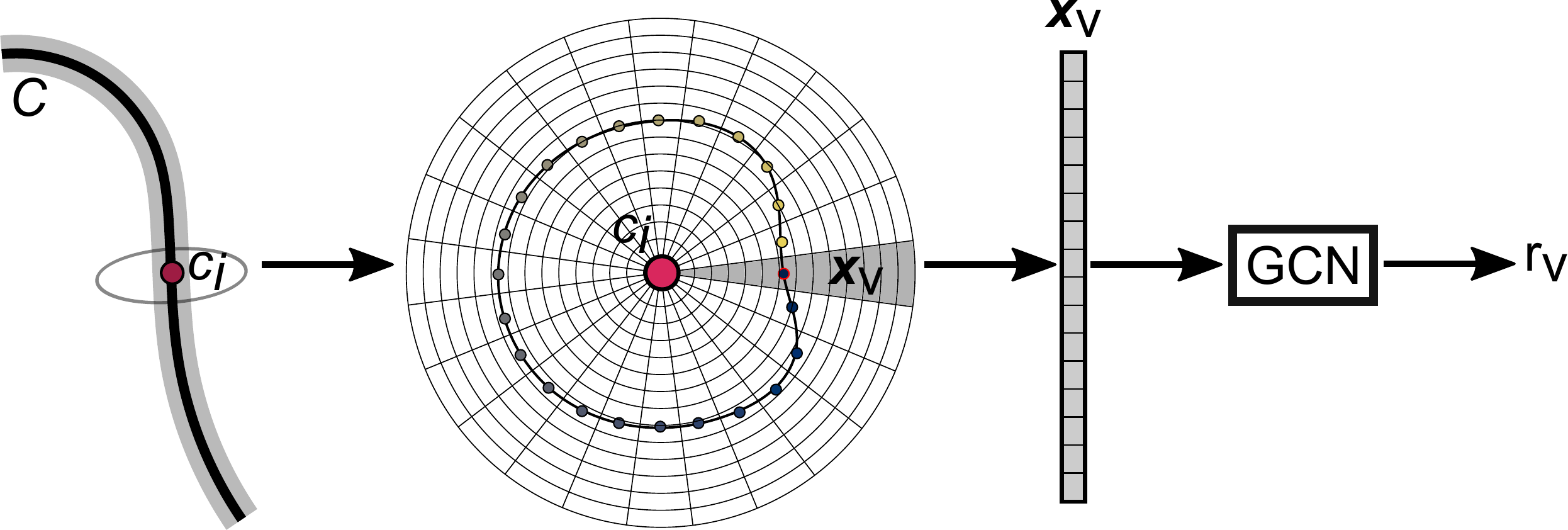}
    \caption{Schematic overview of the proposed method. Given a coronary centerline $C$, rays are cast equiangularly and orthogonally to the centerline at each point $c_i$. Each ray angle corresponds to a vertex $v$ in the luminal surface mesh of which the exact spatial location is determined by $c_i$, an angle $\phi_v$, and the distance $r_v$. Image information along $\mathbf{x}_v$ is used by a GCN that combines this information with that of neighboring vertices to predict $r_v$.}
    \label{fig:1}
\end{figure}

\section{Methods}
We propose to use GCNs to obtain a coronary artery surface mesh based on a CCTA image and an (automatically extracted) coronary artery centerline. Fig. \ref{fig:1} provides an overview of the proposed method. 

\subsection{Surface mesh}
\label{sec:surfacemesh}
The mesh delineating the vessel wall surface is represented by a graph $\mathcal{G(V,E)}$, with vertices $\mathcal{V}$  and edges $\mathcal{E}$. We assume that the vessel wall is a deformable tube in 3D Euclidean space with known centerline $C$. We constrain the exact spatial location of each vertex to be dependent on only one parameter $r_v$. For each centerline point $c_i$, we determine a 2D cross-sectional plane orthogonal to the centerline direction. Within this cross-sectional plane, equiangularly spaced vertices define a cross-section of the surface mesh. Hence, within the 2D plane, the location of each vertex is defined in polar coordinates $(\phi_v, r_v)$, where $\phi_v$ is fixed. The parameter $r_v$ is defined as the distance to the centerline point $c_i$, and it is this value that we predict using regression. 

The combination of all vertices in all cross-sectional planes forms the polygonal lumen surface mesh $\mathcal{G}$. Edges are added between neighboring vertices in cross-sectional planes, and between vertices at the same angle $\phi_v$ in adjacent cross-sectional planes. Four such edges define a quadrilateral mesh face, and each quadrilateral face is further split into two triangular faces. The structure of $\mathcal{G}$ is fixed for a vessel or vessel segment and a GCN is trained to predict the value $r_v$ for each vertex based on information from the image $\mathcal{X}$ encoded in a input vector $\mathbf{x}_v$. In all our experiments, for a vertex $(\phi_v, r_v)$ this input vector contains the image values along a ray cast from the center $c_i$ of the cross-sectional plane at the angle $\phi_v$. 

\begin{figure}[t!]
    \centering
    \includegraphics[width=0.38\linewidth]{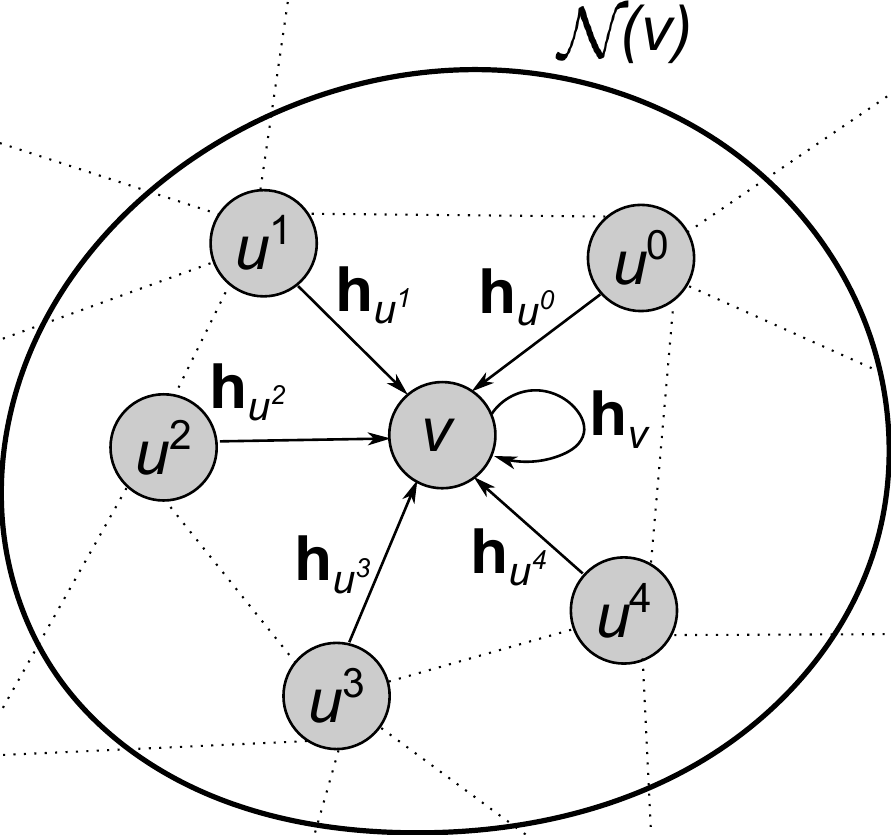}
    \caption{A GCN layer in the proposed method. The graph convolutional network operates on the vertices of the surface mesh and iteratively aggregates information from a vertex $v$ and its neighbors in $\mathcal{N}(v)$ in $\mathbf{h}_v^k$, where $\mathbf{h}_v^0=\mathbf{x}_v$.}
    \label{fig:2}
\end{figure}

\subsection{GCN architecture}
We use a graph convolutional network to predict -- for each node in the graph -- the value of the parameter $r_v$ given the input vector $\mathbf{x}_v$.  The GCN consists of layers that aggregate information from neighboring nodes (Fig. \ref{fig:2}). By concatenating several such layers, information from a growing neighborhood of nodes in the graph is combined.  We use the element-wise mean aggregator as proposed in GraphSAGE \cite{Hami17}. In the $k$-th GCN layer, we aggregate features of vertices $u$ in the neighborhood of vertex $v$, including $v$ itself, to obtain a feature vector $\mathbf{h}_v^k$ for each node. We define $\mathbf{h}_v^0=\mathbf{x}_v$, i.e. the input to the first GCN layer is the input vector derived from the image $\mathcal{X}$. The aggregation rule is

\begin{equation}
    \mathbf{h}_v^k = \sigma\left(\mathbf{W} \cdot \text{MEAN}\left(\{\mathbf{h}_v^{k-1}\} \cup \{\mathbf{h}_u^{k-1}, \forall u \in \mathcal{N}(v)\}\right)\right),
    \label{eq:gcn}
\end{equation}

where $\sigma$ is an activation function, in our case ReLU. The matrix $\mathbf{W}$ contains trainable weights that determine how features are combined between layers. Different than in GraphSAGE, we do not sample neighbors but use the full available neighborhood. The neighborhood is defined by the edges in $\mathcal{E}$, which are fixed for a given vessel segment or segment.

We use a network with five GCN layers, so that predictions for individual vertices are based on vertices that are at most five steps away on the mesh. The first layer has 32 input features in $\mathbf{x}_v$, while the hidden GCN layers each have 64 nodes. The output layer has 1 node to predict $r_v$. Dropout is used with $p=0.5$. Given that we train the GCN using individual samples and not with mini-batches, the GCN does not contain batch normalization layers. In total, the GCN network contains 14,567 trainable parameters. 

The GCN is trained using a training set consisting of coronary artery segments, each of which is represented by a graph $\mathcal{G}_i$, image information $\mathcal{X}_i$ and reference values $r_v$ for all vertices $v \in \mathcal{V}$ in $\mathcal{G}_i$. In each iteration, one segment is randomly selected and the following loss is computed 

\begin{equation}
    \mathcal{L} = \frac{1}{|\mathcal{V}|}\sum_{v \in \mathcal{V}}|r_v^3 - f(\mathbf{x}_v)^3|,
    \label{eq:loss}
\end{equation}

where $\mathcal{V}$ is the set of vertices in $\mathcal{G}_i$ and $f(\mathbf{x}_v)$ is the distance to the centerline predicted by the GCN. Distance values are cubed in the loss function to better guarantee correspondence between automatically segmented and reference vessel volumes. We use no additional regularization on the values of $r_v$. After training, the trained GCN can be applied to a new input graph and image data $(\mathcal{G}', \mathcal{X}')$. 
 
 \begin{table}[t]
\centering
\caption{Quantitative results on the training set of the Coronary Artery Stenoses Detection and Quantification Evaluation Framework. Dice similarity coefficient (DSC), mean surface distance (MSD) and Hausdorff distance (HD) values are shown for human observers as reported in \cite{Kiri13}, previously proposed methods, the proposed graph convolutional network (GCN), and a multi-layer perceptron (MLP).} 
\resizebox{\textwidth}{!}{
\begin{tabular}{l|cc|cc|cc}
       & \multicolumn{2}{c}{DSC} & \multicolumn{2}{c}{MSD (mm)} & \multicolumn{2}{c}{HD (mm)} \\
Method & Healthy    & Diseased    & Healthy       & Diseased      & Healthy      & Diseased      \\ \hline \hline
Expert 1 & 0.74 & 0.79 & 0.26 & 0.26 & 3.61 & 3.29 \\
Expert 2 & 0.66 & 0.73 & 0.25 & 0.31 & 3.00 & 2.70 \\
Expert 3 & 0.80 & 0.76 & 0.23 & 0.24 & 3.25 & 3.07 \\ \hline \hline
Lugauer et al. \cite{Luga14}~ & 0.77 & 0.75 & 0.32 & 0.27 & 2.79 & 1.96 \\
Freiman et al. \cite{Frei17} & 0.69 & 0.74 & 0.49 & 0.28 & 1.69 & 1.22 \\
Mohr et al. \cite{Kiri13} & 0.75 & 0.73 & 0.45 & 0.29 & 3.73 & 1.87 \\
Graph convolutional network (GCN) & 0.75 & 0.73 & 0.25 & 0.28 &1.53 & 1.86 \\ 
Multi-layer perceptron (MLP) & 0.67 & 0.69 & 0.32 & 0.31 & 1.59 & 1.84 \\
\end{tabular}
}
\label{tab:1}
\end{table}

\section{Experiments and Results}
The method was implemented in Python using PyTorch and the Deep Graph Library (DGL)\footnote{\url{https://www.dgl.ai/}}. We performed leave-one-patient-out cross-validation experiments on the training set of the Coronary Artery Stenoses Detection and Quantification Evaluation Framework. The networks were trained using  annotations by all three observers. Coronary centerlines were automatically extracted using the method proposed in \cite{Wolt19} and resampled to 0.5 mm resolution. In all our experiments, we defined 24 discrete values for vertex angles $\phi$ between rays (Fig. \ref{fig:1}) and the resolution of input signals $\mathbf{x}_v$ was 0.1 mm.  Each ray consisted of 32 input features for a circular field of view of 3.2 mm. Image values were clipped at 0 and 1000 HU to normalize for surrounding low-density tissue, air and hyperdense calcifications. We observed that this circumvents the need for explicit calcium removal steps such as proposed in \cite{Luga14}. 

Networks were trained end-to-end and parameters were optimized using the Adam optimizer with a learning rate of 0.001. We trained each network for 50,000 iterations. In each iteration, one training sample $(\mathcal{G}_i, \mathcal{X}_i)$ was randomly selected and the loss in Eq. \ref{eq:loss} was evaluated. To improve training stability, gradients were accumulated for 10 iterations, after which the loss was back-propagated and parameters were updated. Experiments were performed using an NVIDIA Titan X GPU with 12GB of memory. Training of one GCN took around 20 minutes, while segmentation of all coronary arteries in a CCTA image took around 45 seconds, including fully automatic coronary centerline extraction \cite{Wolt19}. 

\begin{figure}[t!]
    \centering
    \includegraphics[width=0.95\textwidth]{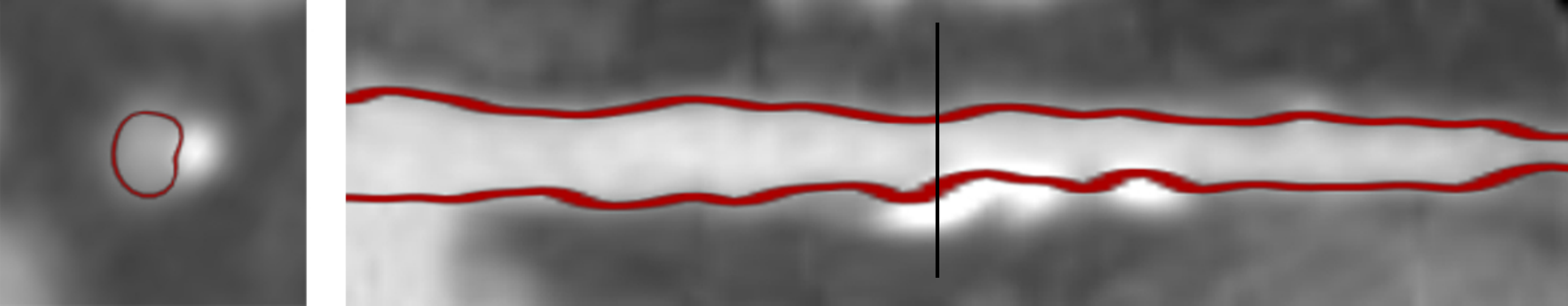}
    \caption{Example segmentation, showing how the GCN has learned to exclude coronary artery calcium from its segmentation. The figure on the left shows the cross-sectional plane indicated by a black line on the straightened multi-planar reconstruction on the right.}
    \label{fig:3}
\end{figure}

We evaluated the quality of the obtained segmentations using the online platform of the Coronary Artery Stenoses Detection and Quantification Evaluation Framework \cite{Kiri13}. Dice similarity coefficients (DSC) were determined based on 2D cross-sectional segmentations obtained along the centerline. In addition, the mean surface distance (MSD) and Hausdorff distance (HD) were computed between automatic and reference meshes. Table \ref{tab:1} lists results the three expert observers in \cite{Kiri13}, state-of-the-art methods evaluated on the same 78 segments in 18 CCTA images, and the GCN. Results are shown separately for healthy and diseased segments. The results indicate that the GCN obtains results that are comparable to those obtained by other methods in terms of all three metrics, and that like other automatic methods, the GCN outperforms human experts in terms of Hausdorff distance. Fig. \ref{fig:3} shows an example segmentation in an artery with coronary artery disease. The segmentation follows the inner vessel wall and excludes coronary artery calcification.

\begin{figure}[t!]
    \centering
    \subfloat[MLP]{
        \includegraphics[width=0.48\linewidth]{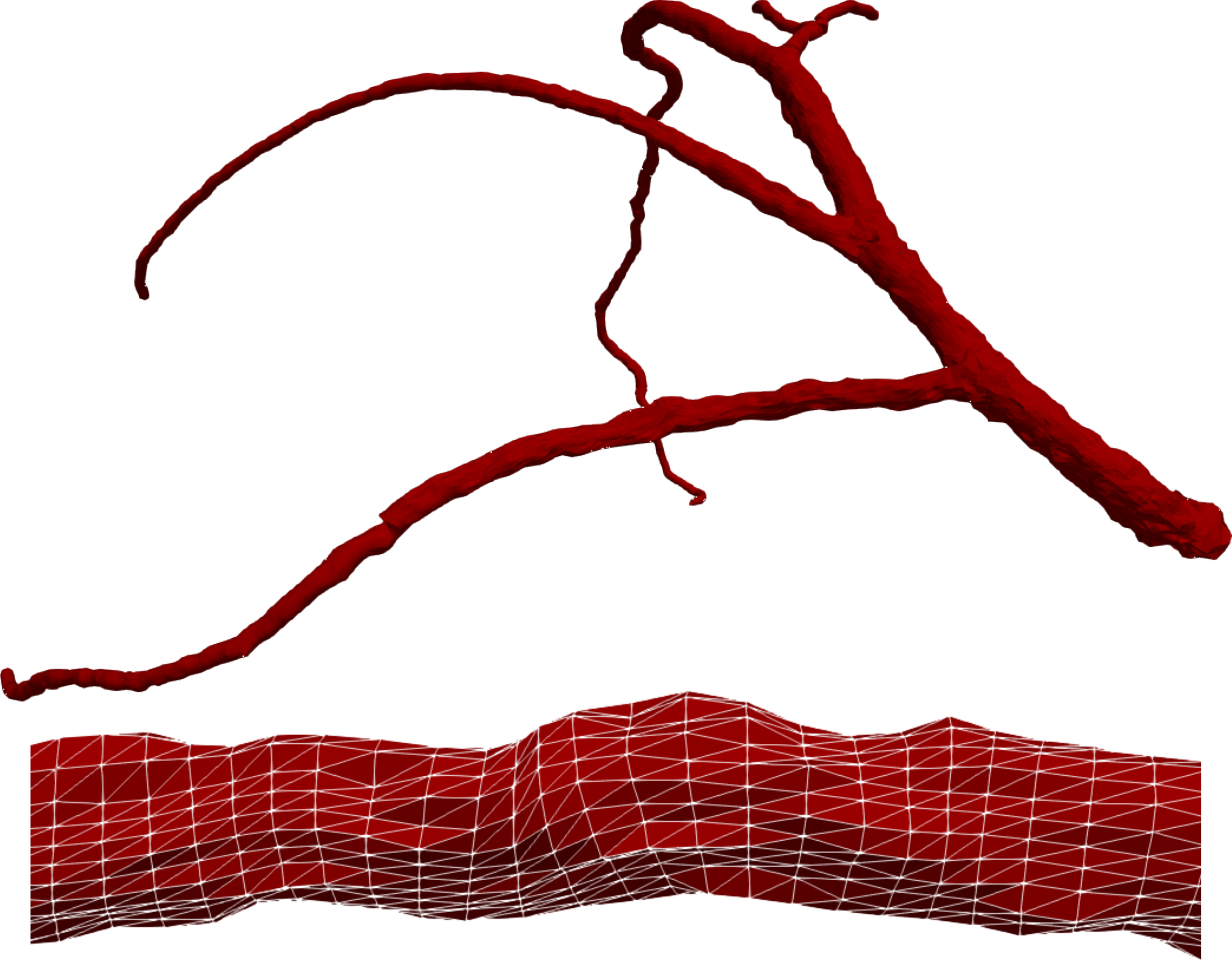} 
        \label{subfig:4a}
    }
    \hfill
    \subfloat[GCN]{
        \includegraphics[width=0.48\linewidth]{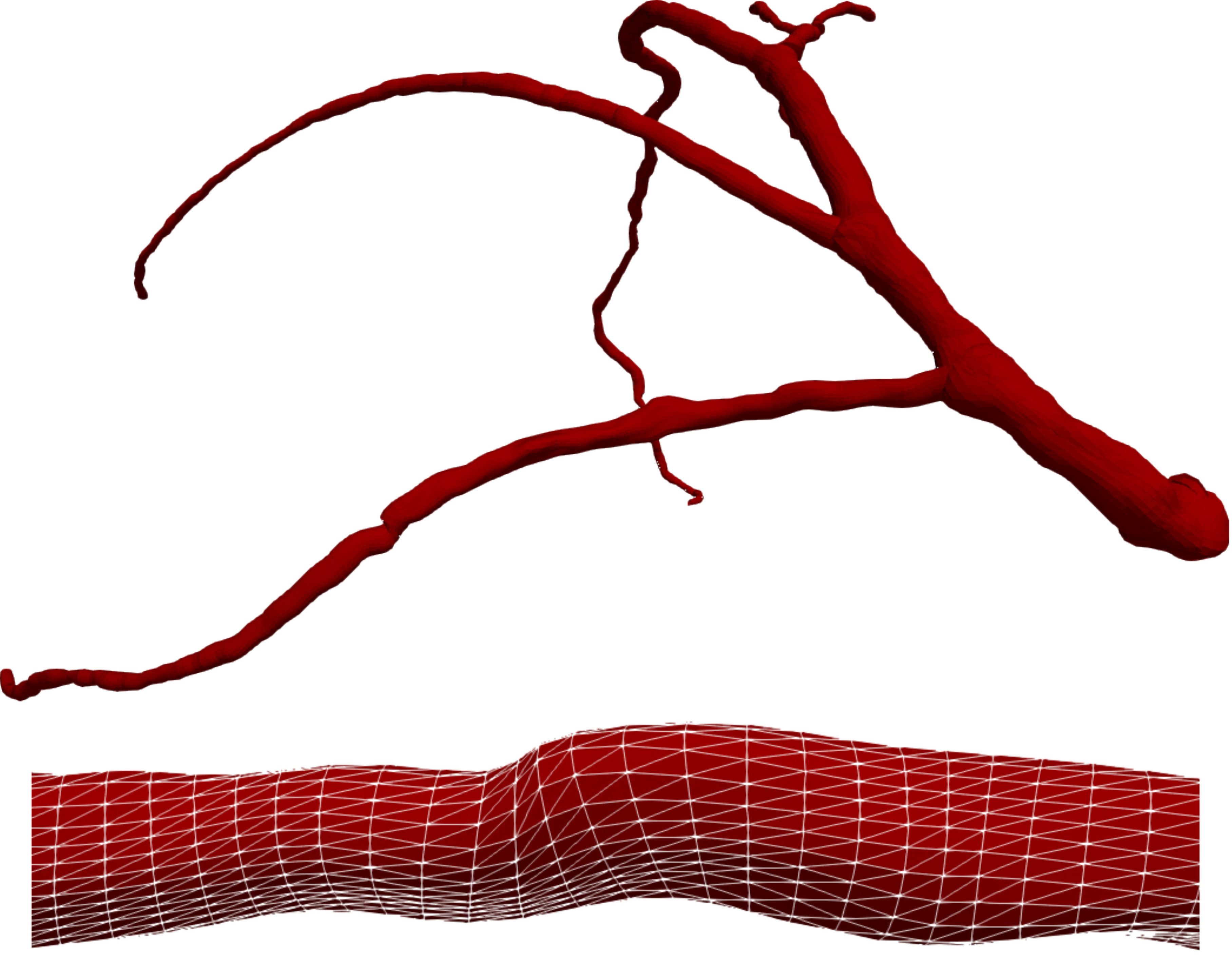} 
        \label{subfig:4b}
    }
    \caption{Left coronary artery tree segmented using \protect\subref{subfig:4a} multi-layer perceptron (MLP) and \protect\subref{subfig:4b} graph convolutional network (GCN). By including information from neighboring vertices, the GCN leads to a smoother surface than the MLP.}
    \label{fig:4}
\end{figure}

In addition, we performed experiments to determine the value of GCN layers over traditional fully-connected layers. We removed the ability to propagate values between neighboring nodes on the surface mesh from the layer in Eq. \ref{eq:gcn}. The result is a standard fully-connected layer with $\mathbf{h}_v^k = \sigma\left(\mathbf{W} \cdot \mathbf{h}_v^{k-1}\right)$. Like in Eq. \ref{eq:gcn}, this layer only has trainable parameters in $\mathbf{W}$. We built a network which contained only fully-connected layers, i.e. a multi-layer perceptron (MLP) and compared this network to the GCN. Both networks contained 14,567 trainable parameters and both networks were trained in the same manner. Fig. \ref{fig:4} shows an example of a left coronary artery tree segmented using the MLP or the GCN. By combining information from neighboring vertices on the surface, the GCN intrinsically leads to smoother segmentations than the MLP. Quantitative results in Table \ref{tab:1} show that the inclusion of GCN layers in the network architecture leads to substantially higher overlap (DSC) and better accuracy (MSD).

\section{Discussion and Conclusions}
We have proposed a method based on graph convolutional networks for coronary artery segmentation in cardiac CT angiography. Experiments using a publicly available framework showed competitive performance on segmentation of individual artery segments. To the best of our knowledge, this work is among the first applications of GCNs in medical image analysis, following recent works on airway extraction in chest CT \cite{Selv18} and cortical segmentation in brain MR \cite{Cucu18}. The method obtained accurate results in both healthy and diseased vessels. Given that segmentations meet requirements of down-stream applications, i.e. contiguous segmentations with sub-voxel accuracy, future work could include validation of the obtained segmentations for anatomical and functional stenosis detection. 

We found that when using the same experimental settings on the same data, the inclusion of GCN-layers substantially improved segmentation performance over a baseline network using only fully-connected layers. This indicates that the GCN successfully uses information from a local neighborhood on the surface mesh to predict the location of an individual mesh vertex. Moreover, the method does not require post-processing steps such as conditional random fields, or additional regularization terms to smooth the obtained surface meshes, but directly generates a smooth surface mesh. 

A limitation of the mean aggregator used in this work is its invariance to the spatial relation between a vertex and its neighbors. In future work, we will investigate the use of edge features in addition to the node features used in this work. Alternatively, information from neighboring vertices could be combined in different ways, such as convolution in cross-sectional planes. An additional limitation that this work shares with other methods for coronary lumen segmentation \cite{Luga14,Frei17,Lee19} is the dependence on a coronary artery centerline. In preliminary experiments, we found that the exact location of this coronary artery centerline can lead to noticeable differences in segmentation accuracy. We will further investigate this in future work. 

In conclusion, GCN-based segmentation of coronary arteries in CCTA is feasible and accurate.

\subsubsection{Acknowledgements}
P15-26, Project 2, Dutch Technology Foundation with participation of Philips Healthcare.

\bibliographystyle{splncs03}
\bibliography{bibliography}

\end{document}